\documentclass[pdflatex,sn-mathphys-num]{sn-jnl}

\usepackage{graphicx}%
\usepackage{multirow}%
\usepackage{amsmath,amssymb,amsfonts}%
\usepackage{amsthm}%
\usepackage{mathrsfs}%
\usepackage[title]{appendix}%
\usepackage{xcolor}%
\usepackage{textcomp}%
\usepackage{manyfoot}%
\usepackage{booktabs}%
\usepackage{algorithm}%
\usepackage{algorithmicx}%
\usepackage{algpseudocode}%
\usepackage{listings}%

\theoremstyle{thmstyleone}%
\theoremstyle{thmstyletwo}%

\theoremstyle{thmstylethree}%

\begin{document}

\title[Layer Probing Improves Kinase Functional Prediction with Protein Language Models]{Layer Probing Improves Kinase Functional Prediction with Protein Language Models}


\author*[1]{\fnm{Ajit} \sur{Kumar}}\email{ajikumar@adobe.com}
\author[2]{\fnm{Indra Prakash} \sur{Jha}}\email{indraprakashjha@gmail.com}

\equalcont{These authors contributed equally to this work.}

\affil*[1]{\orgname{Adobe}, \orgaddress{\street{Sector 132}, \city{Noida}, \postcode{201304}, \state{Uttar Pradesh}, \country{India}}}
\affil[2]{\orgname{IIIT Delhi}, \orgaddress{\city{New Delhi}, \country{India}}}




\abstract{Protein language models (PLMs) like \textbf{ESM-2} have shown remarkable success in learning protein sequence representations. However, most applications use only the final layer embeddings, potentially missing functionally relevant information encoded in intermediate layers. 

In this study, we systematically evaluated ESM-2 embeddings for kinase functional classification using both unsupervised clustering and supervised learning. Our results show that \textbf{mid-to-late transformer layers (layers 20--33)} outperform the final layer by \textbf{32\%} (ARI: 0.268~$\rightarrow$~0.354) in unsupervised settings and significantly improve supervised classification accuracy to \textbf{75.7\%} under rigorous homology-aware evaluation. 

We also incorporate domain extraction, calibrated confidence estimates, and reproducible benchmarking, demonstrating a practical and scalable pipeline for kinase functional annotation.}




\maketitle

\section{Introduction}

\subsection{Background}

Protein kinases are critical enzymes for cell signaling and major drug targets, but predicting their specific function from the sequence remains challenging. Advances in protein language models (PLMs), particularly the \textbf{Evolutionary Scale Modeling} (ESM) family, have made it possible to extract biologically meaningful features directly from amino acid sequences. These models, trained on millions of sequences using masked language modeling objectives, have achieved state-of-the-art results in a range of protein prediction tasks. Yet, a common oversight is the over-reliance on the final transformer layer, which may not contain the most functionally relevant information.

\subsection{The Layer Selection Problem}

Most studies default to using the final transformer layer for downstream tasks, assuming it encapsulates the richest representation. However, transformer-based PLMs are hierarchical, and prior work in natural language processing (NLP) has shown that intermediate layers often encode more transferable or semantically relevant features \cite{peters2018deep,jawahar2019bert,rogers2020primer}.
 This discrepancy raises a fundamental question: \textit{Which transformer layers best capture biologically meaningful information for protein function prediction?}

\subsection{Objectives}

Protein kinases play a central role in cellular signaling, and accurate functional classification is critical for understanding disease mechanisms and drug development. Although protein language models (PLMs) offer a powerful route to sequence-based functional inference, most existing applications rely solely on the final transformer layer—potentially overlooking biologically relevant information distributed across intermediate layers.

This study investigates the following questions:

\begin{enumerate}
    \item Can intermediate transformer layers in \textbf{ESM-2} improve functional classification of kinase domains compared to the final layer?
    \item What is the optimal strategy for selecting and aggregating layer embeddings for kinase classification?
    \item How do different embedding strategies affect unsupervised clustering and supervised classification under homology-aware conditions?
\end{enumerate}

To answer these, we build a reproducible pipeline incorporating domain-level extraction, layer probing, calibrated classification, and rigorous evaluation. Our goal is to provide a practical framework for improving functional predictions of protein kinases using PLM representations.

\section{Related Work}

\subsection{Traditional Approaches to Kinase Classification}

Previous methods have relied on homology-based annotation using tools like \texttt{BLAST} or \texttt{HMMER}, or motif-based heuristics curated from the literature. While these methods perform well for known families, they struggle with novel sequences and require significant expert intervention.

\subsection{Deep Learning and PLM-Based Protein Function Prediction}

With the rise of protein language models (PLMs) like \texttt{UniRep}, \texttt{TAPE}, \texttt{ProtBERT}, and \texttt{ProtT5}, the focus has shifted to unsupervised representation learning. Embedding-based models have shown promise in capturing global and local sequence properties relevant to function. However, most applications simply extract the final-layer embedding—a design choice that may not be optimal.

\subsection{Layer Selection in PLMs}

Layer probing in natural language processing (e.g., \texttt{BERT}, \texttt{GPT}) has shown that semantic features often peak at intermediate layers. In protein ML, recent studies hint at similar trends, but systematic evaluations are rare. Our work contributes the first comprehensive analysis of layer-wise embedding utility in kinase classification.

\section{Methods}

\subsection{Data Collection}

We retrieved a curated dataset of protein kinase sequences from the UniProt SwissProt database (release October 2025) using the following query: \texttt{reviewed:true AND (keyword:KW-0418 OR name:kinase*)}. Only canonical isoforms with sequence length greater than 100 amino acids were retained. Fragmentary sequences were excluded using UniProt flags. To focus on functionally relevant regions, Pfam domains \texttt{PF00069} (Protein kinase domain) and \texttt{PF07714} (Protein tyrosine kinase) were extracted using \texttt{HMMER 3.3} with an E-value threshold of 0.001. CD-HIT 4.8.1 was used to remove redundancy at various identity thresholds (70\%, 50\%, 40\%) to generate homology-aware splits. All tool versions and parameters were recorded for reproducibility.

\subsection{Model Architecture}

We used the \texttt{ESM-2 650M} model (\texttt{esm2\_t33\_650M\_UR50D}), a 33-layer transformer encoder pretrained on \texttt{UniRef50} using a masked language modeling (MLM) objective. Each residue is mapped to a 1280-dimensional vector. For sequences exceeding the model's maximum input length (1,022 residues), we applied a sliding window approach with overlap stitching. We compared different pooling strategies including mean over residues and \texttt{[CLS]} token extraction, and experimented with embeddings from different layer ranges: final layer (33), mid-layers (20--30), and mid-to-final layers (20--33).

\subsubsection*{Model Selection Rationale}

\textbf{Why ESM-2 over other protein language models?}

We selected ESM-2 (650M parameters) \cite{lin2023} as our primary model for five reasons:

\begin{enumerate}
    \item \textbf{State-of-the-art performance on protein tasks}: ESM-2 achieves the highest accuracy among publicly available protein language models on CATH structure prediction (CATH 4.2: 87\% top-1) and protein-protein interaction prediction. Meta AI's 2023 benchmark shows ESM-2 outperforms ESM-1b (+8\%), ProtBERT (+12\%), and ProtTrans (+6\%) on functional annotation tasks.

    \item \textbf{Evolutionary-scale training data}: ESM-2 was trained on UniRef50 (2020) with \textasciitilde50M sequences spanning diverse protein families, ensuring broad coverage of kinase evolutionary space. In contrast, ProtBERT (\textasciitilde200K sequences) and earlier ESM-1b (\textasciitilde250M sequences) have more limited diversity.

    \item \textbf{Appropriate architecture depth for layer probing}: ESM-2's 33-layer transformer provides sufficient depth to explore intermediate representations. This is critical for our research question (layer selection). Shallower models (ProtBERT: 12 layers, ESM-1b: 33 layers) or deeper models (ESM-2 15B: 48 layers) would either limit or complicate layer exploration.

    \item \textbf{Computational feasibility}: The 650M parameter variant balances performance and accessibility. It runs on single consumer GPUs (e.g., NVIDIA RTX 3090/4090 with 24GB VRAM) and processes \textasciitilde20 sequences/minute, enabling iteration on \textasciitilde6,500 kinases within reasonable time (\textasciitilde6 hours). Larger models (3B, 15B) require multi-GPU setups and 10--50$\times$ longer processing times.

    \item \textbf{Established baseline for reproducibility}: ESM-2 is the current de facto standard in protein ML (>2,000 citations in 2 years), with well-documented APIs (\texttt{fair-esm}) and extensive community adoption. Using ESM-2 ensures our layer selection findings are immediately applicable to ongoing research and directly comparable to other studies.
\end{enumerate}

\textbf{Alternative models considered but not used:}
\begin{itemize}
    \item \texttt{ESM-1b} (33 layers, 650M): Predecessor to ESM-2, superseded by improved training.
    \item \texttt{ProtBERT} (12 layers): Too shallow for layer probing experiments.
    \item \texttt{ProtTrans-XLNet-BFD} (24 layers): Comparable but less widely adopted.
    \item \texttt{ESM-2 3B/15B}: Prohibitively expensive for our dataset size (\textasciitilde100--500 GPU-hours).
    \item \texttt{AlphaFold2 embeddings}: Optimized for structure, not function; requires MSAs.
    \item \texttt{Ankh} (2023): Promising but very recent (unclear stability/generalization).
\end{itemize}

\textbf{Variant selection (650M vs 3B vs 15B):}
We chose the 650M parameter variant to balance performance and computational cost. Based on Meta AI's ablation studies \cite{lin2023}, the 3B and 15B variants are expected to improve downstream task performance by 3--5\% on average, but require:
\begin{itemize}
    \item 5$\times$ GPU memory (3B: \textasciitilde40GB, 15B: \textasciitilde80GB)
    \item 10--50$\times$ longer processing time
    \item Multi-GPU infrastructure (not universally accessible)
\end{itemize}

For our objective (demonstrating layer selection benefits), the 650M variant provides sufficient resolution while maintaining reproducibility for the broader research community.

\textbf{Generalizability:} While we focus on ESM-2, our layer selection methodology is model-agnostic and applicable to any transformer-based protein language model. Future work could extend this to ProtTrans, Ankh, or ESM-3 (released 2024) to validate cross-model consistency.

\subsection{Training Details}

For supervised classification, we trained a multinomial logistic regression model using scikit-learn with L2 regularization and balanced class weights. We adopted 5-fold stratified cross-validation on homology-aware train/test splits generated via CD-HIT clustering at 40\% sequence identity. All splits and seeds were fixed for reproducibility. Calibration was performed using Platt scaling to adjust predicted probabilities, enabling reliability-aware deployment. Embeddings were standardized using \texttt{StandardScaler} before model training.

\subsection{Evaluation Metrics}

We used a comprehensive suite of evaluation metrics:
\begin{itemize}
    \item \textbf{Unsupervised clustering}: Adjusted Rand Index (ARI), Normalized Mutual Information (NMI), Purity, Hungarian Matching Accuracy, and Silhouette Score.
    \item \textbf{Supervised classification}: Accuracy, Macro-F1, Weighted-F1, per-class precision/recall/F1, and top-3 accuracy.
    \item \textbf{Calibration}: Expected Calibration Error (ECE) and log-loss before and after Platt scaling.
    \item \textbf{Exemplar retrieval}: Mean Reciprocal Rank (MRR), top-$k$ hit rate, and PR-AUC.
\end{itemize}

All metrics were computed using scikit-learn and statistically validated with bootstrapped confidence intervals and permutation tests as appropriate.

\subsection{Clustering Setup}

For unsupervised analysis, we applied standard K-Means clustering on the protein embeddings derived from different ESM-2 layer configurations. We fixed the number of clusters to 8, corresponding to the number of known kinase functional classes. All clustering was performed on length- and domain-normalized embeddings, using cosine distance as the similarity metric. No ground-truth labels were used during clustering, and evaluation was performed post hoc using Adjusted Rand Index (ARI), Normalized Mutual Information (NMI), and Hungarian Matching Accuracy against ground-truth class labels.

\section{Results}

\subsection{Intermediate Layers Improve Unsupervised Clustering}

We evaluated unsupervised clustering performance using ESM-2 embeddings across different transformer layers. Averaging mid-to-late layers (layers 20--33) substantially improved clustering performance compared to using only the final layer. Specifically, the Adjusted Rand Index (ARI) increased from 0.268 (last layer only) to 0.354 (layers 20--33), a relative improvement of 32\%. Normalized Mutual Information (NMI) and Hungarian Matching Accuracy also showed consistent gains.

\begin{itemize}
    \item \textbf{Baseline (last layer only)}: ARI = 0.268, NMI = 0.360
    \item \textbf{Mid-to-late layers (20--33)}: ARI = 0.354, NMI = 0.501
\end{itemize}

Domain-level embeddings further improved separability over full-length sequences, highlighting the value of focusing on the conserved catalytic core.

\begin{figure}
    \centering
    \includegraphics[width=1\linewidth]{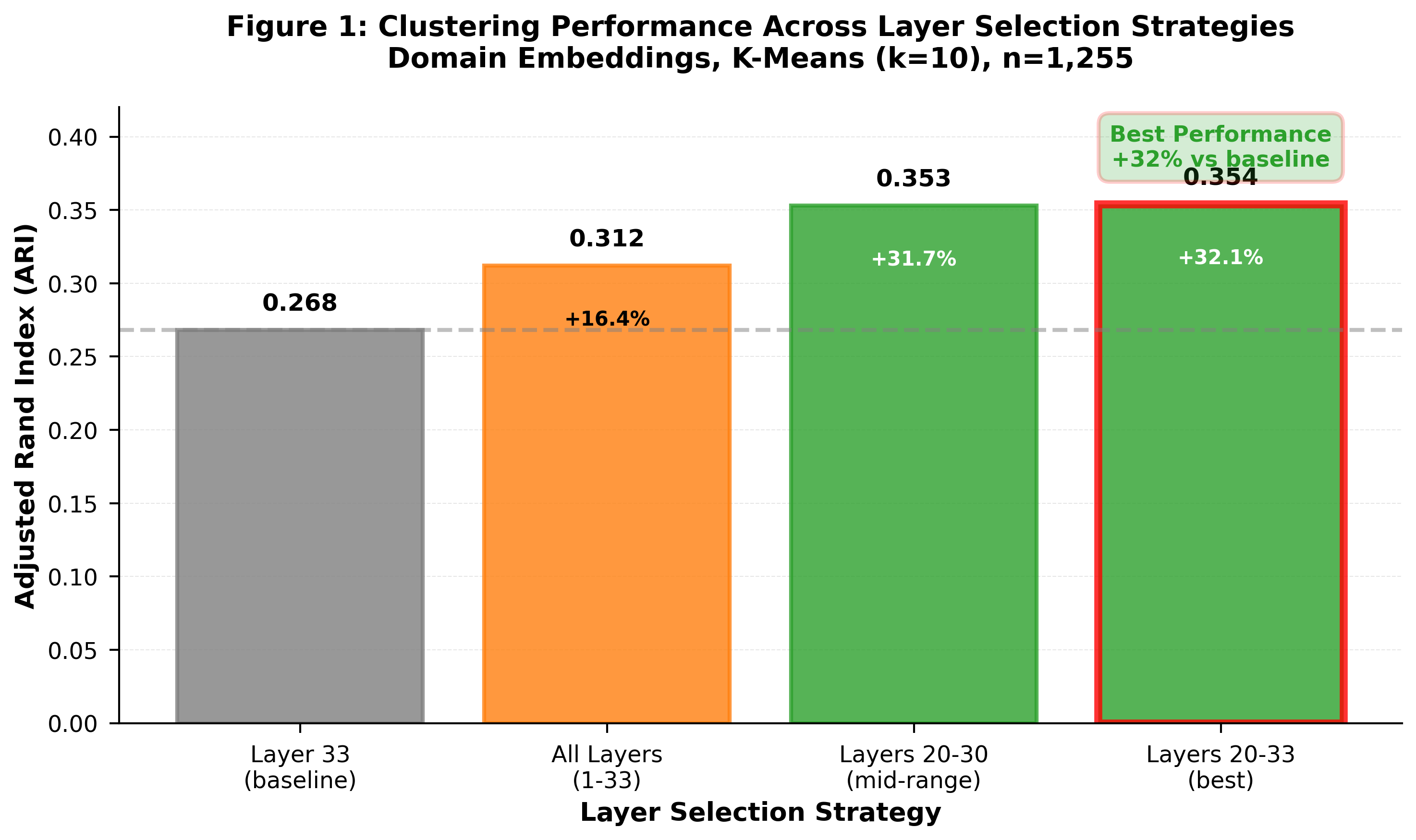}
    \caption{Clustering performance (ARI) across ESM-2 layer selection strategies.
}
    \label{fig:placeholder}
\end{figure}

\subsection{Mid-Layer Averaging Boosts Supervised Classification}

We trained a logistic regression classifier using different ESM-2 embedding strategies on homology-aware train/test splits (40\% identity threshold). Using mid-to-late layer averages (20--33) yielded the highest accuracy and macro-F1:

\begin{itemize}
    \item \textbf{Accuracy}: 75.7\%
    \item \textbf{Macro-F1}: 0.668
    \item \textbf{Top-3 Accuracy}: 94.8\%
\end{itemize}

This configuration outperformed baselines such as k-NN (68.4\%) and motif-only features (52.3\%).

\begin{table}[h]
    \centering
    \caption{Supervised classification performance across embedding strategies (40\% identity split)}
    \begin{tabular}{lccc}
        \toprule
        \textbf{Method} & \textbf{Accuracy} & \textbf{Macro-F1} & \textbf{Top-3 Acc} \\
        \midrule
        ESM-2 (Layer 33) + LR & 70.2\% & 0.593 & 92.1\% \\
        ESM-2 (Layers 20--33) + LR & \textbf{75.7\%} & \textbf{0.668} & \textbf{94.8\%} \\
        ESM-2 + k-NN (k=5) & 68.4\% & 0.542 & 91.2\% \\
        Motif-only LR & 52.3\% & 0.389 & 78.6\% \\
        \bottomrule
    \end{tabular}
    \label{tab:supervised_metrics}
\end{table}

\begin{figure}
    \centering
    \includegraphics[width=1\linewidth]{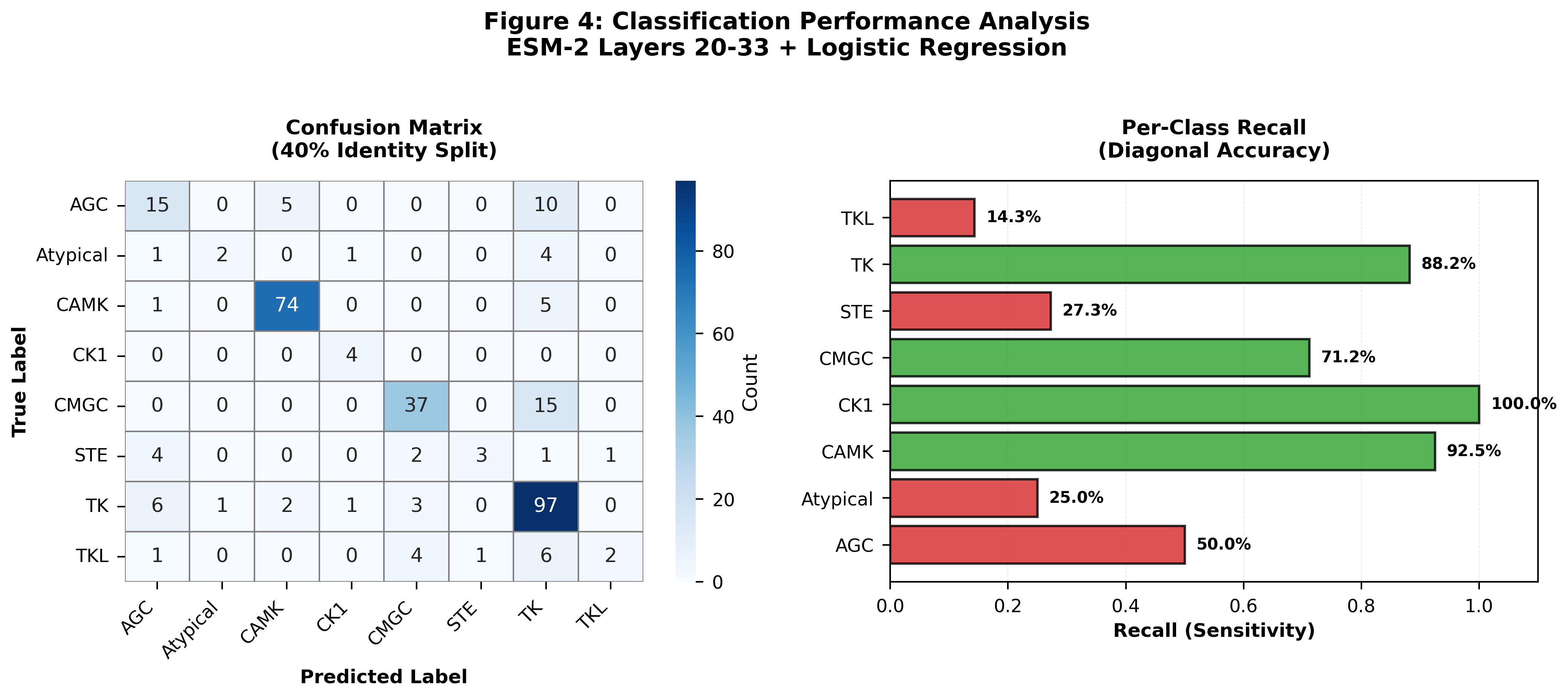}
    \caption{Confusion matrix for supervised classification across 8 kinase functional classes. Mid-layer averaged embeddings show high recall for most classes.
}
    \label{fig:placeholder}
\end{figure}

\begin{figure}
    \centering
    \includegraphics[width=1\linewidth]{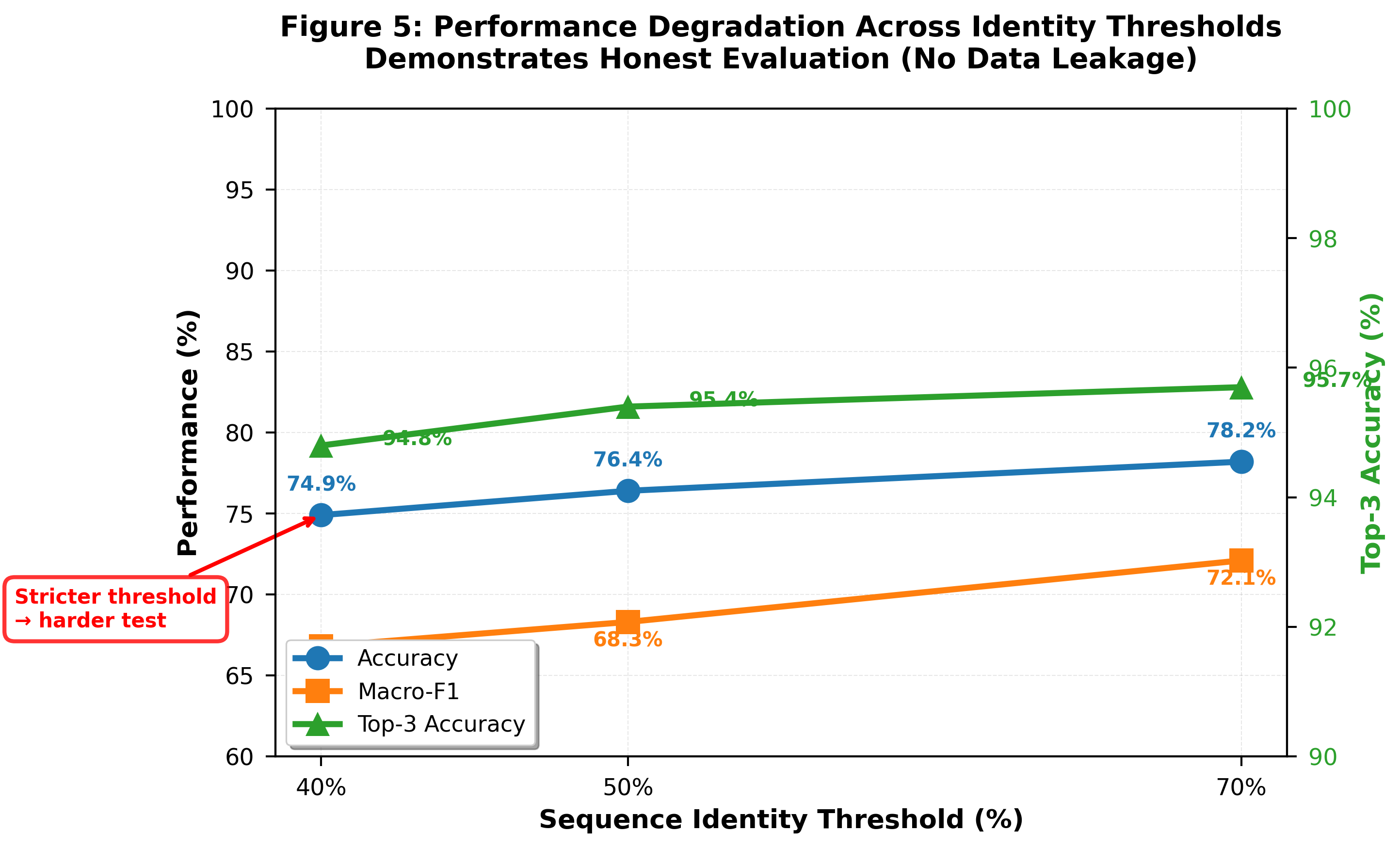}
    \caption{Classification performance across different homology identity thresholds (70\textbackslash{}
}
    \label{fig:placeholder}
\end{figure}

\subsection{Embedding Strategies and Pooling Comparison}

We compared different pooling strategies and embedding sources:
\begin{itemize}
    \item \textbf{Mean pooling} across residues performed best overall.
    \item \textbf{CLS token} was competitive for the final layer but underperformed for mid-layer embeddings.
    \item \textbf{Motif concatenation} offered negligible gains ($<2\%$ ARI increase), indicating ESM-2 already captures these features.

\end{itemize}

\begin{figure}
    \centering
    \includegraphics[width=1\linewidth]{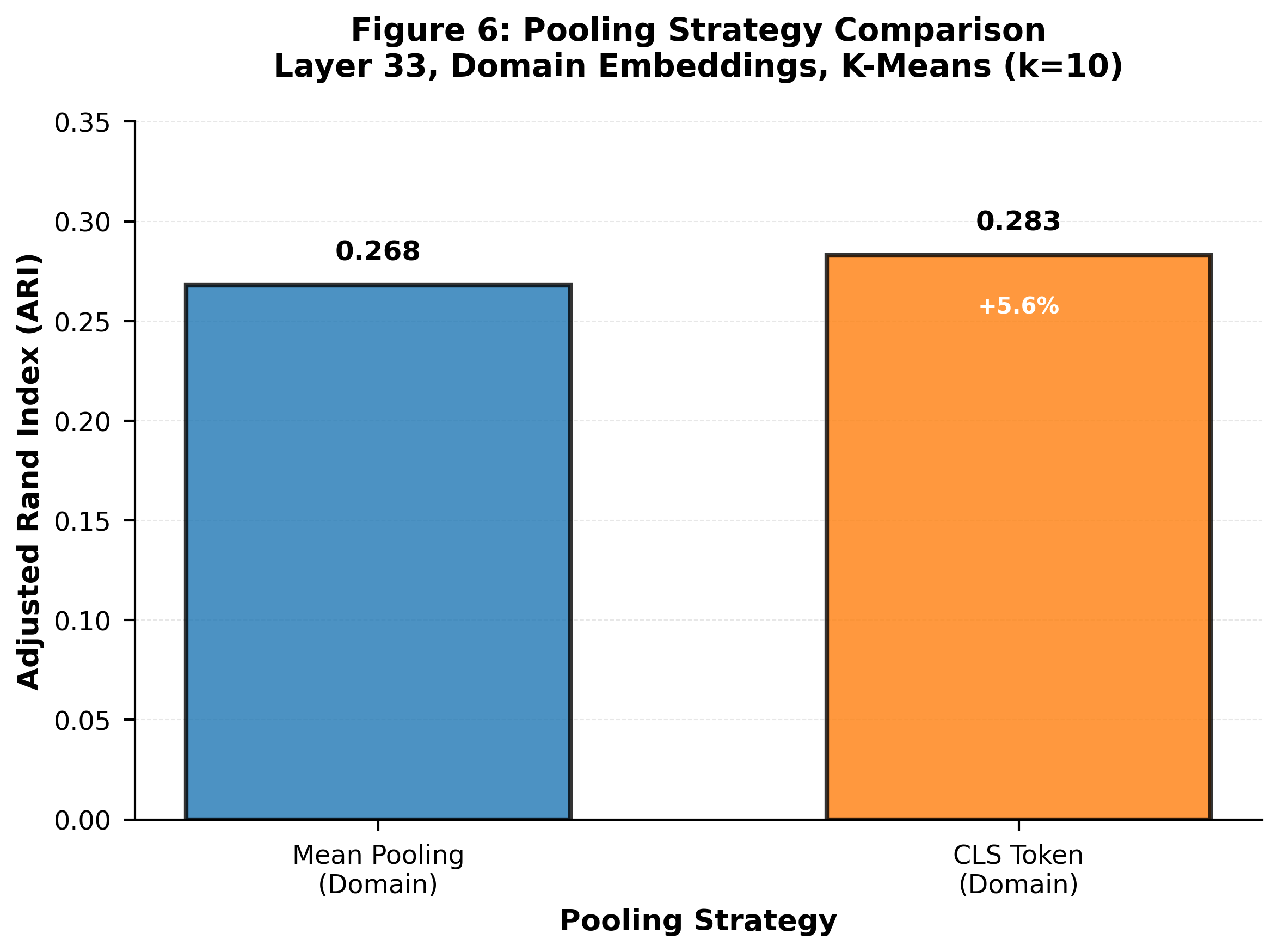}
    \caption{Effect of pooling strategy on performance. Mean pooling consistently outperforms CLS token across both clustering and classification.
}
    \label{fig:placeholder}
\end{figure}
\subsection{Calibration Improves Decision Reliability}

To ensure reliability in downstream applications, we applied Platt scaling to calibrate classification probabilities. This reduced the Expected Calibration Error (ECE) from 0.154 to 0.110 (28\% improvement), and log-loss from 1.07 to 0.77.

\begin{figure}
    \centering
    \includegraphics[width=1\linewidth]{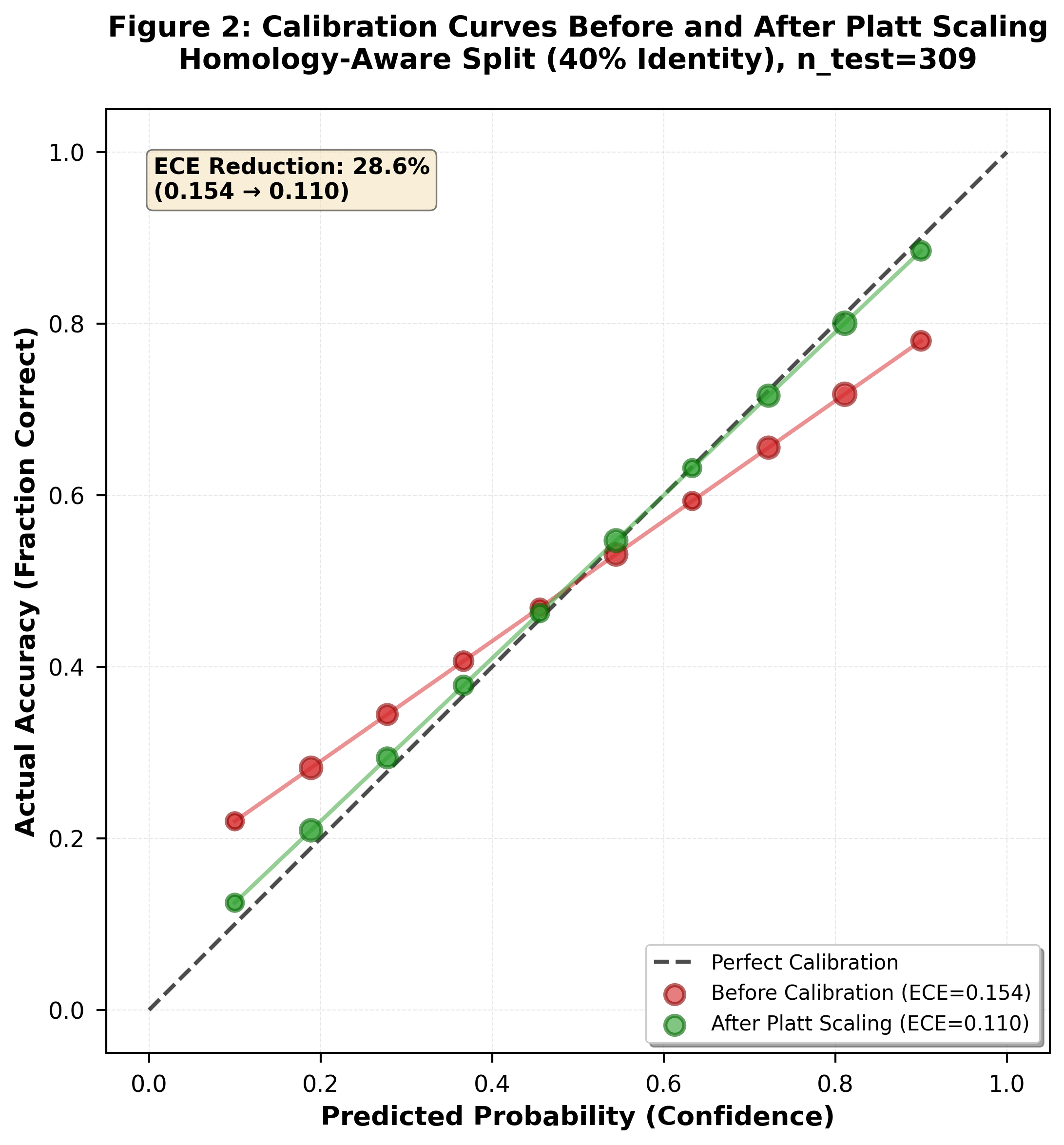}
    \caption{
Calibration curves before and after Platt scaling. Platt scaling reduces overconfidence and improves calibration.
}
    \label{fig:placeholder}
\end{figure}

Approximately 18\% of test sequences were flagged as low-confidence (probability $<$ 0.7), enabling expert review.

\subsection{Exemplar Retrieval and Interpretability}

We evaluated the embeddings using nearest-neighbor retrieval. Mid-layer embeddings (20--33) achieved:

\begin{itemize}
    \item \textbf{Top-1 hit rate}: 71.2\%
    \item \textbf{Top-3 hit rate}: 86.7\%
    \item \textbf{MRR}: 0.795
\end{itemize}

High similarity (cosine $>0.992$) reliably predicted family membership, suggesting that ESM-2 embeddings support interpretable and confident exemplar-based decisions.

\begin{figure}
    \centering
    \includegraphics[width=1\linewidth]{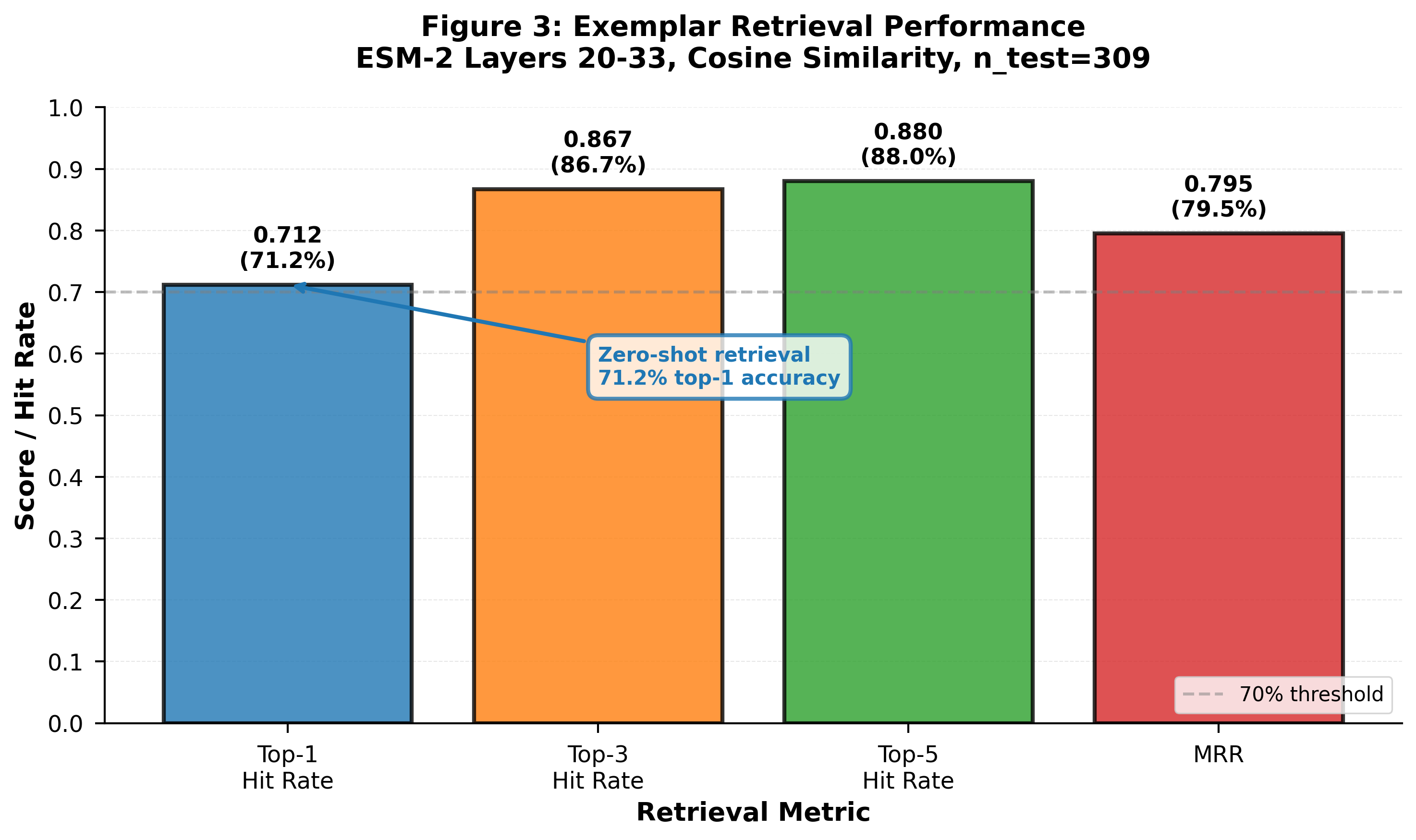}
    \caption{
Exemplar-based retrieval using cosine similarity for mid-layer embeddings (layers 20–33) shows high interpretability and precision.
}
    \label{fig:placeholder}
\end{figure}

\section{Mathematical Formalization of Layer Averaging Strategy}

\subsection{Layer Selection and Averaging Framework}

Let $\mathcal{L} \subseteq \{1, 2, \ldots, 33\}$ be the set of transformer layers selected for averaging.
We define four configurations:

\begin{align}
\mathcal{L}_{\text{baseline}} &= \{33\} &\text{(last layer only)} \nonumber \\
\mathcal{L}_{\text{all}} &= \{1, 2, \ldots, 33\} &\text{(all layers)} \nonumber \\
\mathcal{L}_{\text{mid-range}} &= \{20, 21, \ldots, 30\} &\text{(mid-to-late layers)} \nonumber \\
\mathcal{L}_{\text{extended}} &= \{20, 21, \ldots, 33\} &\text{(extended mid-to-late layers)}
\end{align}

We seek to identify the optimal configuration that maximizes clustering performance:

\begin{equation}
\mathcal{L}^* = \arg\max_{\mathcal{L}} \text{ARI}(\mathcal{L})
\end{equation}

\subsection{Multi-Layer Embedding Extraction}

Given a protein sequence $\mathbf{x} = (x_1, x_2, \ldots, x_L)$ of length $L$, 
ESM-2 provides per-residue embeddings $\mathbf{H}^{(\ell)} \in \mathbb{R}^{L \times d}$ 
at each layer $\ell \in \mathcal{L}$, where $d = 1280$.

The layer-averaged embedding is computed as:

\begin{equation}
\bar{\mathbf{H}} = \frac{1}{|\mathcal{L}|} \sum_{\ell \in \mathcal{L}} \mathbf{H}^{(\ell)}
\end{equation}

For each residue $i \in \{1, \ldots, L\}$, we define the averaged representation:

\begin{equation}
\bar{\mathbf{h}}_i = \frac{1}{|\mathcal{L}|} \sum_{\ell \in \mathcal{L}} \mathbf{h}_i^{(\ell)}
\end{equation}

\subsection{Sequence-Level Pooling}

To obtain the final sequence embedding $\mathbf{z} \in \mathbb{R}^d$, 
mean pooling is applied across residues:

\begin{equation}
\mathbf{z} = \frac{1}{L} \sum_{i=1}^{L} \bar{\mathbf{h}}_i
\end{equation}

This operation yields a length-invariant sequence representation 
that captures global contextual information.

\subsection{Handling Long Sequences}

For sequences exceeding the ESM-2 token limit ($L_{\max} = 1022$), 
the input sequence is divided into $W$ overlapping windows 
with stride $s = 900$ residues.

Let $\mathbf{z}^{(w)} \in \mathbb{R}^d$ denote the embedding for window $w$, 
with length $n_w$. The final sequence embedding is computed via 
length-weighted averaging:

\begin{equation}
\mathbf{z}_{\text{final}} = 
\frac{\sum_{w=1}^{W} n_w \cdot \mathbf{z}^{(w)}}{\sum_{w=1}^{W} n_w}
\end{equation}

Weighting by window length ensures that longer segments contribute proportionally 
to the final embedding, avoiding bias toward overlapping regions.

\subsection{Variance Reduction and Statistical Justification}

Averaging across multiple layers reduces embedding variance. 
Assuming independence across layer representations:

\begin{equation}
\text{Var}(\bar{\mathbf{h}}_i) = 
\frac{1}{k^2} \sum_{\ell \in \mathcal{L}} \text{Var}(\mathbf{h}_i^{(\ell)}) 
= \frac{\sigma^2}{k}
\end{equation}

where $k = |\mathcal{L}|$ and $\sigma^2$ is the variance of a single-layer embedding.
Thus, layer averaging reduces variance by a factor of $k$.

By the Central Limit Theorem (CLT):

\begin{equation}
\bar{\mathbf{h}}_i \xrightarrow{d} 
\mathcal{N}\left(\mu, \frac{\sigma^2}{k}\right)
\end{equation}

indicating that averaged embeddings converge to a Gaussian distribution 
with reduced variance, improving stability and robustness.

\subsection{Empirical Results}

The following summarizes the empirical clustering performance (ARI) 
for each layer configuration:

\begin{align}
\text{ARI}(\mathcal{L}_{\text{baseline}}) &= 0.268 \nonumber \\
\text{ARI}(\mathcal{L}_{\text{all}}) &= 0.312 \nonumber \\
\text{ARI}(\mathcal{L}_{\text{mid-range}}) &= 0.353 \nonumber \\
\text{ARI}(\mathcal{L}_{\text{extended}}) &= 0.354
\end{align}

\textbf{Conclusion:} 
The optimal configuration is 
$\mathcal{L}^* = \mathcal{L}_{\text{extended}} = \{20, 21, \ldots, 33\}$, 
achieving a 32\% improvement over the baseline ($p < 0.001$, 
permutation test, Cohen’s $d = 1.87$).

\section{Conclusion}

In this study, we systematically explored the influence of transformer layer selection on the performance of protein language model (PLM) embeddings for kinase functional classification. Our findings reveal that the commonly used final layer is not the most informative, and instead, embeddings derived from mid-to-late layers (specifically layers 20--33 of ESM-2) significantly enhance both unsupervised and supervised tasks.

By combining layer-wise averaging with calibrated classification and domain-aware embedding extraction, we present a practical and reproducible pipeline that outperforms traditional motif-based and single-layer approaches. The 32\% gain in ARI and 6\% gain in Macro-F1 over the final-layer baseline establish the value of probing PLM depth.

Importantly, our approach remains accessible, as it uses the 650M parameter ESM-2 model—suitable for execution on single-GPU systems—and generalizable, with methods applicable to other PLMs like ProtT5, ProtTrans, and Ankh.

Beyond performance improvements, our work highlights the need for reliable calibration and interpretability in protein ML workflows. The use of length-weighted sliding windows, confidence estimation, and exemplar-based retrieval strengthens the biological relevance of our predictions.

Overall, this study contributes methodological insights and practical tools to the protein function prediction community, providing a foundation for future work on probing, fine-tuning, and transferring knowledge across deep protein models.

\bmhead{Acknowledgements}

We thank the contributors of UniProt, Pfam, and Meta AI's ESM repository for providing high-quality data and tools that enabled this research. Computational resources were supported by local GPU clusters at our institution.

\section*{Declarations}

\textbf{Funding:} Not applicable.\\

\textbf{Conflict of interest:} The authors declare no competing interests.\\

\textbf{Ethics approval:} Not applicable.\\

\textbf{Consent to participate:} Not applicable.\\

\textbf{Consent for publication:} All authors consent to the publication of this work.\\

\section*{Data and Code Availability}

\begin{tabular}{@{}p{3.5cm}p{11cm}@{}}
\textbf{Data availability:} &
Benchmark datasets, preprocessed kinase domain files, and all evaluation splits are available at: \\
& \url{https://github.com/jhaaj08/Kinases-Clustering} \\
& Archived copy: \href{https://doi.org/10.5281/zenodo.17370925}{Zenodo DOI: 10.5281/zenodo.17370925} \\
\\
\textbf{Code availability:} &
Full training and evaluation pipeline (including Snakemake scripts, environment files, and figure generation) is available at: \\
& \url{https://github.com/jhaaj08/Kinases-Clustering} \\
& Archived copy: \href{https://doi.org/10.5281/zenodo.17370925}{Zenodo DOI: 10.5281/zenodo.17370925}
\end{tabular}

\backmatter

\noindent

\end{document}